\documentstyle[multicol,aps,epsf,pra]{revtex}
\begin{document}
\draft

\title{Preparation of Nonclassical States in Cavities with a 
Moving Mirror}
\author{S.\ Bose, K.\ Jacobs and P.\ L.\ Knight}
\address{Optics Section, The Blackett Laboratory, Imperial 
College, London SW7 2BZ, England}

\maketitle
\begin{abstract} We describe how a quantum system composed of a 
cavity field interacting with a movable mirror can be utilized to 
generate a large variety of nonclassical states of both the cavity 
field and the mirror. First we consider state preparation of the 
cavity field. The system  dynamics will prepare  a single mode of 
the cavity field in a multicomponent Schr\"{o}dinger cat state, in 
a similar manner to that in a Kerr medium. In addition, when two or more
cavity modes interact with the mirror, they may be prepared in an 
entangled state which may be regarded as a multimode generalisation of
even and odd coherent states. We show 
also that near-number states of a single mode may be prepared by 
performing a measurement of the position of the mirror. Secondly 
we consider state preparation of the mirror, and show that this 
macroscopic object may be placed in a Scr\"{o}dinger cat state 
by a quadrature measurement of the light field. In addition we 
examine the effect of the damping of the motion of the  mirror 
on the field states inside the cavity, and compare this with 
the effect of cavity field damping.
\end{abstract}


\begin{multicols}{2}

\section{Introduction}
Recently, Mancini {\em et.al.}~\cite{Tombesi} have shown how 
a cavity with a movable mirror (treated as a quantum harmonic 
oscillator) can be used to synthesize Schr\"{o}dinger cat 
states~\cite{Buzek,Monroe} of the cavity field. In fact, as we 
will see here, this system can lead to the production of an 
extensive class of novel states of the cavity field, including 
entangled states of two or more cavity modes.  One of the aims 
of this paper is to describe how these can be achieved. 
Furthermore, in Ref.\cite{Tombesi} only the effect of photon 
leakage from the cavity was considered as a relevant source of 
decoherence. We examine the opposite extreme, namely when the 
decoherence induced in the field due to the damping of the 
mirror's motion is the dominant source of decoherence and photon 
leakage is almost absent. We calculate the effect of this type 
of decoherence on the states of the cavity field and explicitly 
demonstrate that, due to the oscillatory nature of the system, 
the decoherence rate is much lower than expected. Moreover, we 
show that apart from trying to improve the mirror's isolation, 
increasing its frequency also helps to reduce the rate of the 
type of decoherence considered here. This fact comes as a benefit 
in the generation of atleast one type of nonclassical state of 
the cavity field. These results definitely brighten the prospects 
of observing nonclassical states of the field inside the cavity 
given that damping of the mirror's motion is inevitable. It is 
also shown that even when the effect of mirror damping on the 
cavity field is dominant, it does not destroy all the 
nonclassical features of the cavity field.

In Ref.\cite{Tombesi}, the main focus was the generation of 
nonclassical states of the cavity field. We give here a 
thorough treatment of the quantum dynamics of the  mirror 
as well. In particular, we point out ways in which this mirror 
motion can also be put in a nonclassical state. Of course, if 
these are to be observed, then very good isolation of our 
system from the environment is necessary. However, the very 
fact that this system allows in principle the production of 
nonclassical states of a macroscopic object such as the 
mirror (after sufficient isolation) should be interesting in 
itself. 

\section{Dynamics of the undamped system}
Numerous authors have previously treated the system of a cavity 
field and a movable mirror, both quantum 
mechanically~\cite{Tombesi,Kurt} and 
classically~\cite{Meyes,Walther}, and both 
theoretically~\cite{Tombesi,Kurt,Meyes} and experimentally 
\cite{Walther,Hein}. We here assume the movable mirror to 
be a quantum harmonic oscillator with frequency $\omega_m$ 
and annihilation operator denoted by $b$, interacting with 
a cavity field mode of frequency $\omega_0$ and annihilation 
operator denoted by $a$. The relevant Hamiltonian~\cite{Tombesi} 
is
\begin{equation}
\label{g}
  H = \hbar\omega_0~a^\dagger a ~+~\hbar\omega_m~b^\dagger 
b~-~\hbar g~a^\dagger a (b+b^\dagger)
\end{equation}
where
\begin{equation}
\label{gpar}
  g~=~\frac{\omega_0}{L} ~\sqrt\frac{\hbar}{2m\omega_m} ~,
\end{equation} 
and  $L$ and $m$ are the length of the cavity and mass of 
the movable mirror respectively.

The time evolution operator corresponding to the above 
Hamiltonian was evaluated by Mancini {\em et al.}~\cite{Tombesi} 
(for completeness we give a proof in Appendix A) and is given by
\end{multicols}
\vspace{-0.5cm}
\noindent\rule{0.5\textwidth}{0.4pt}\rule{0.4pt}{\baselineskip}
\begin{equation}
\label{ev}
  U(t)=\mbox{exp}\left[-i r a^\dagger a ~t\right]
       \mbox{exp}\left[i k^2 (a ^\dagger a)^2 (t-\sin{t})\right]
       \mbox{exp}\left[k a^\dagger a(\eta b^\dagger-
\eta^*b)\right]        
       \mbox{exp}\left[-i b^\dagger b t\right] ~,
\end{equation}
\begin{multicols}{2}
where $\eta=(1-e^{-it})$, $k=g/\omega_m$ is the scaled 
coupling parameter, $r=\omega_0/\omega_m$, and $t$ represents 
a scaled time, being the actual time multiplied by $\omega_m$. 
We note that values for $k$ of the order of unity are 
experimentaly feasible ($\omega_0\sim 10^{16}\mbox{ s}^{-1}, 
\omega_m\sim 1\mbox{ Khz}, L\sim 1\mbox{ m}, m\sim 10\mbox{ mg}$)~\cite{Tombesi,Meyes,Walther}. We see from the above 
equation that there is now an explicit Kerr-like term in $U(t)$, 
so that physically one might expect the cavity field to have an 
evolution similar to that in a Kerr like nonlinearity. In view 
of the large variety of nonclassical states that can be produced 
by a Kerr medium~\cite{Tanas,Gar,Ag}, this system clearly offers 
prospects for the production of nonclassical states of 
light~\cite{Tombesi,Hein}. Indeed, it is known that, like the Kerr 
nonlinearity, this system of a moving-mirror cavity also exhibits 
optical bistability~\cite{Meyes,Walther}.

Let us assume that initially both the mirror and the cavity field 
are in coherent states. To see that this is reasonable we first 
note that the long-time steady state of the cavity mode is the 
vacuum, and that the most stable pointer state~\cite{point} for 
the mirror (being a single harmonic oscillator) is a coherent 
state~\cite{Z}. The steady state therefore consists of the cavity 
mode in the vacuum state and the mirror in a coherent state. 
The cavity mode can now be placed in a non-vacuum coherent state 
by driving it with a coherent input field on a time-scale which 
is much shorter than the time-scale of the mirror motion. Thus 
we write the initial state at time $t=0$ as  
\begin{equation} 
  |\Psi(0)\rangle = |\alpha\rangle_{\mbox{\scriptsize c}} \otimes |\beta\rangle_{\mbox{\scriptsize m}}
\end{equation}
where $|\alpha\rangle_{\mbox{\scriptsize c}}$ and 
$|\beta\rangle_{\mbox{\scriptsize m}}$ are initial coherent 
states of the field and the mirror respectively.
\begin{figure}[h]
\begin{center} 
\leavevmode 
\epsfxsize=8cm 
\epsfbox{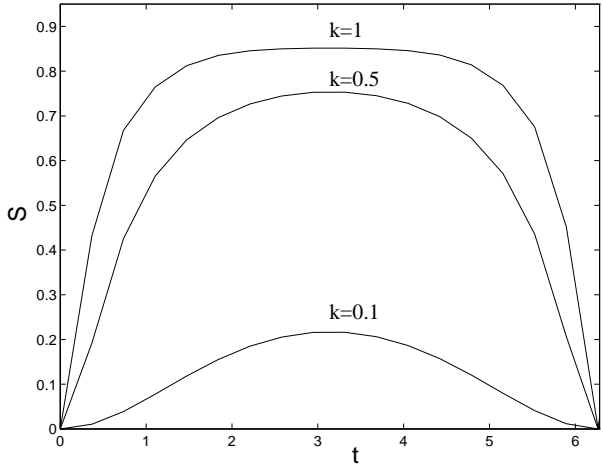}
\caption{\narrowtext The linear entropy, $S$, of the mirror 
state, in the absence of any damping, is plotted here as a 
function of time, and for various values of the scaled coupling 
parameter $k$. We have taken the initial coherent state 
amplitude of the cavity mode to be $\alpha=2$, and that of 
the mirror to be $\beta=2$. The entanglement of the mirror 
with cavity field increases from a value of zero to a maximum 
at $t=\pi$ and subsequently falls back to zero at $t=2\pi$. 
Both the scaled time $t$ and the linear entropy $S$ are 
dimensionless quantities.}
\end{center}
\end{figure}

The time evolution of the system in the interaction picture 
(that is, omitting the free evolution of the field) leads to 
a state at time $t$ given by
\begin{eqnarray}
\label{ro1}
  |\Psi(t)\rangle~=~e^\frac{-|\alpha|^2}{2}\sum_{n=0}^\infty
\frac{\alpha^n}{\sqrt{n!}}e^{ik^2 n^2 (t-\sin{t})}|n
\rangle_{\mbox{\scriptsize c}} \otimes |\phi_n (t)\rangle_{\mbox{\scriptsize m}}
\end{eqnarray}
where $|n\rangle_{\mbox{\scriptsize c}}$ denotes a Fock 
state of the cavity field with eigenvalue $n$, and the 
$|\phi_n (t)\rangle_{\mbox{\scriptsize m}}$ are coherent 
states of the mirror given by
\begin{equation}
\label{phi1}
  |\phi_n (t)\rangle_{\mbox{\scriptsize m}} =|~\beta e^{-it} 
+ kn(1-e^{-i t})~\rangle_{\mbox{\scriptsize m}} ~.
\end{equation}
Two features of the mirror dynamics emerge from the above 
equations:

(1)~ After a time $t=2\pi$ the mirror returns to its original 
state. At all times between $t=0$ and $t=2\pi$ the mirror 
state is entangled with the field state with the entanglement 
being maximum when $t=\pi$. The time dependence of the 
entanglement is shown in Fig.(1) as a plot of the purity, or 
``linear entropy''~\cite{en} which is given by $(1-Tr(
\rho_M(t)^2))$ where $\rho_M(t)$ is the reduced density 
matrix of the mirror.

(2)~ The mirror, always being in a mixture of coherent states 
during its evolution, is described by entirely positive Wigner 
functions and therefore has a fully classical dynamics as 
illustrated in Fig.(2). In this figure, and all the figures 
showing the Wigner function of the mirror, $x$ shall stand 
for $b + b^\dagger$ and $y$ for $-i(b-b^\dagger)$. The mirror 
undergoes an oscillation of a different amplitude (but at the 
same frequency) for each number state of the field. The net 
effect is a kind of {\em breathing} of the mirror state; the 
mirror state undergoes an oscillatory increase and decrease of 
its position and momentum spread. However, this is very different 
from the well known breathing of squeezed states, as the mirror 
state is a mixture of coherent states.
\begin{figure} 

\begin{center} 
\leavevmode 
\epsfxsize=8cm 
\epsfbox{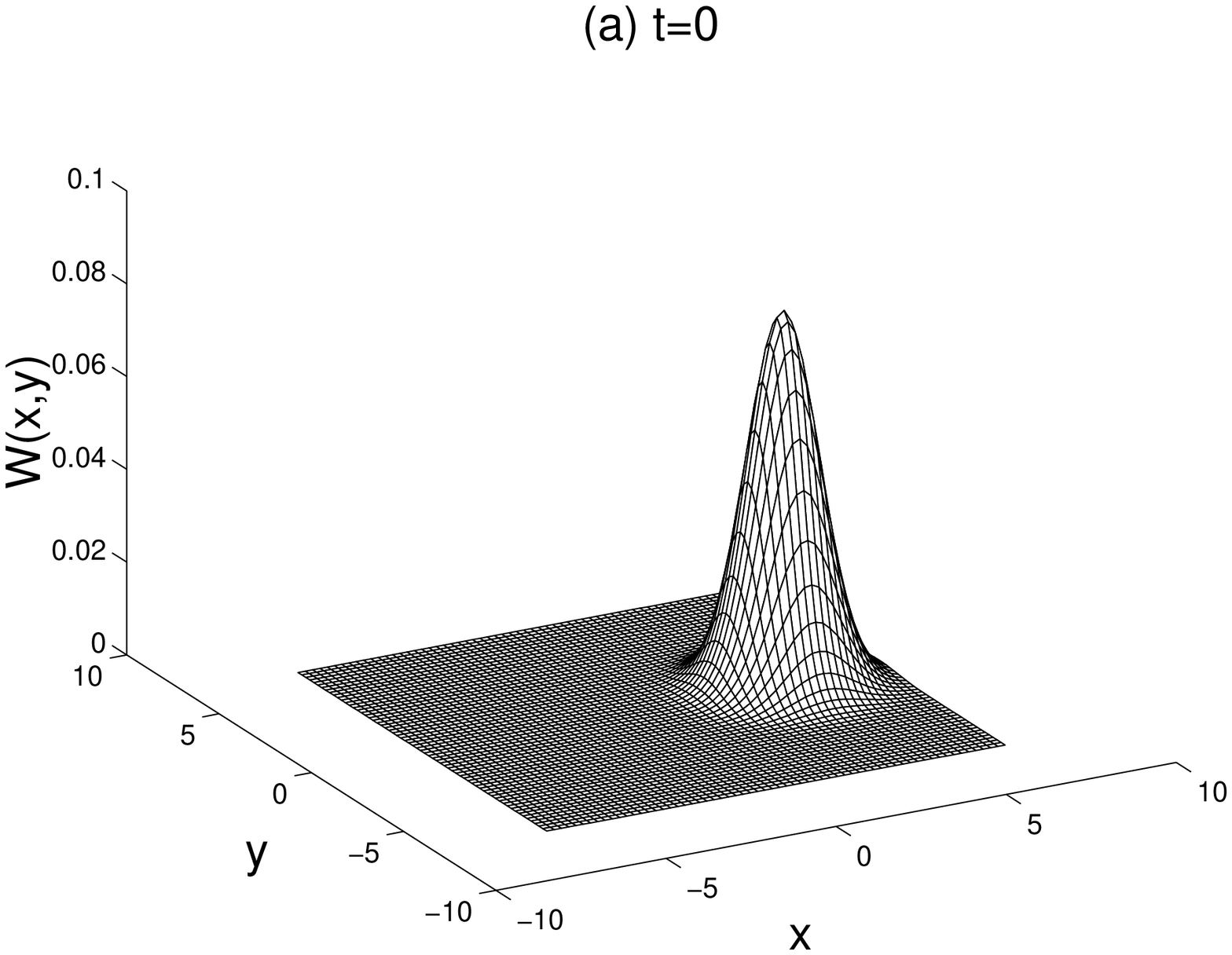}
\label{setup2} 
\end{center} 
 
\begin{center} 
\leavevmode 
\epsfxsize=8cm 
\epsfbox{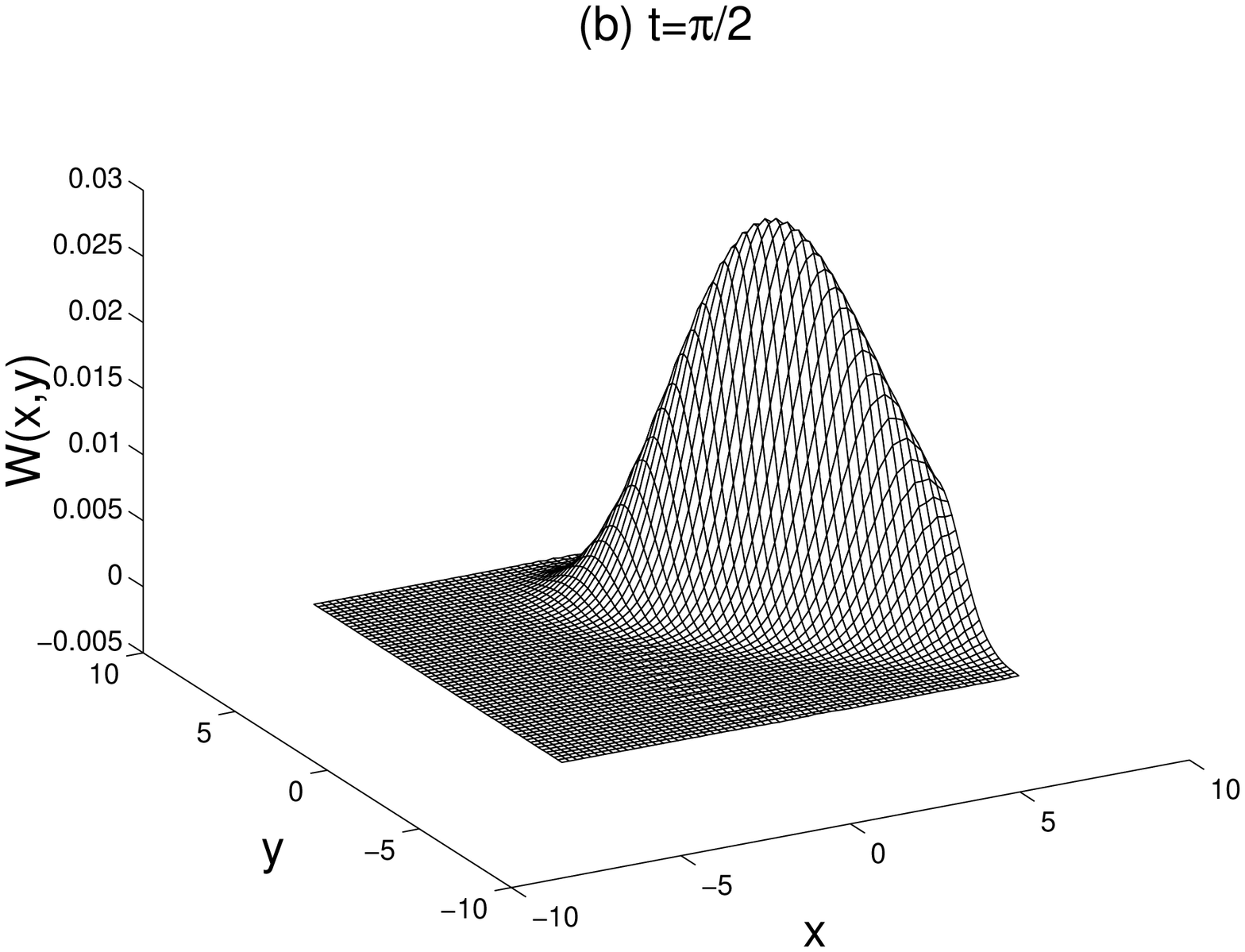}
\end{center}
\begin{center} 
\leavevmode 
\epsfxsize=8cm 
\epsfbox{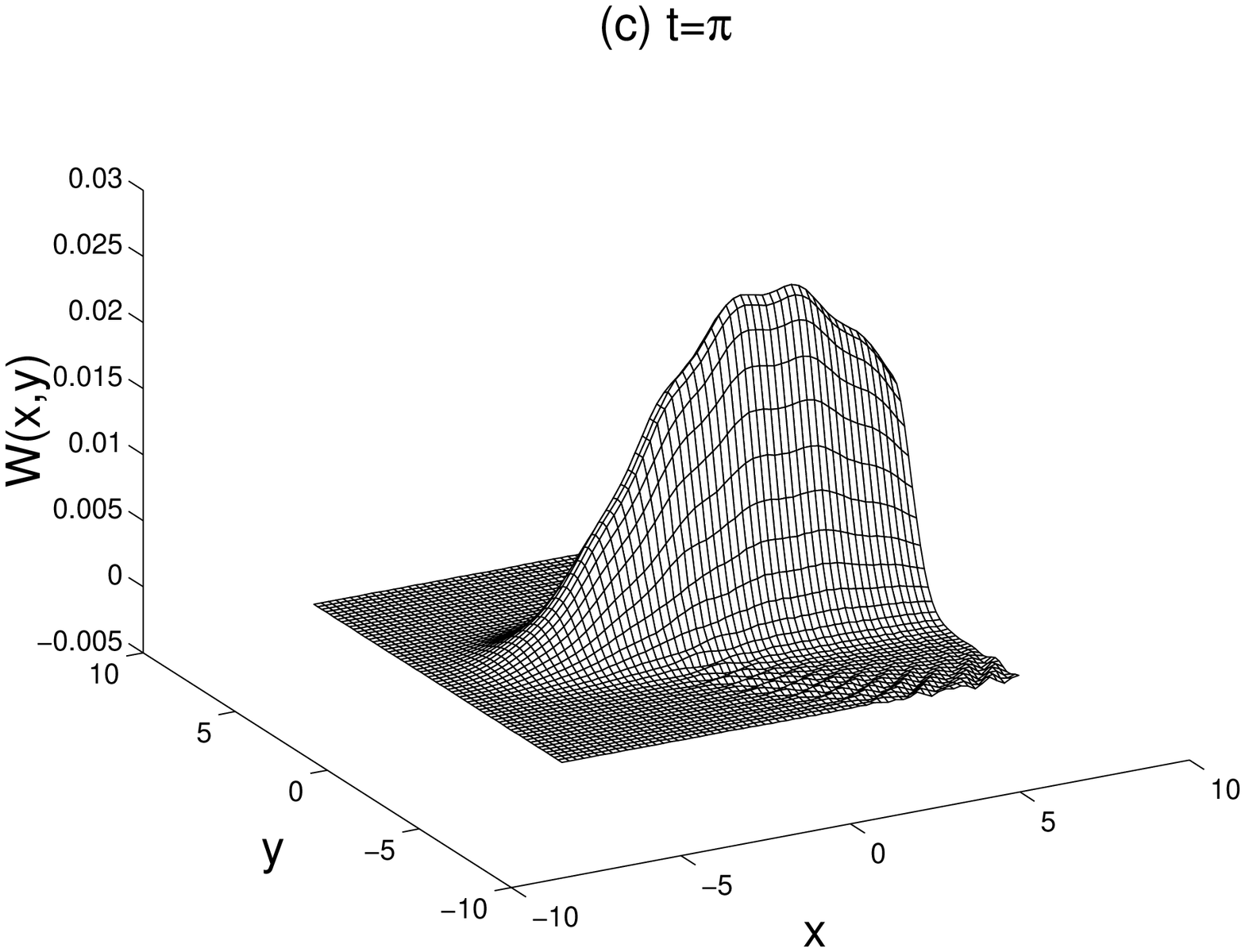}
\end{center} 
\begin{center} 
\leavevmode 
\epsfxsize=8cm 
\epsfbox{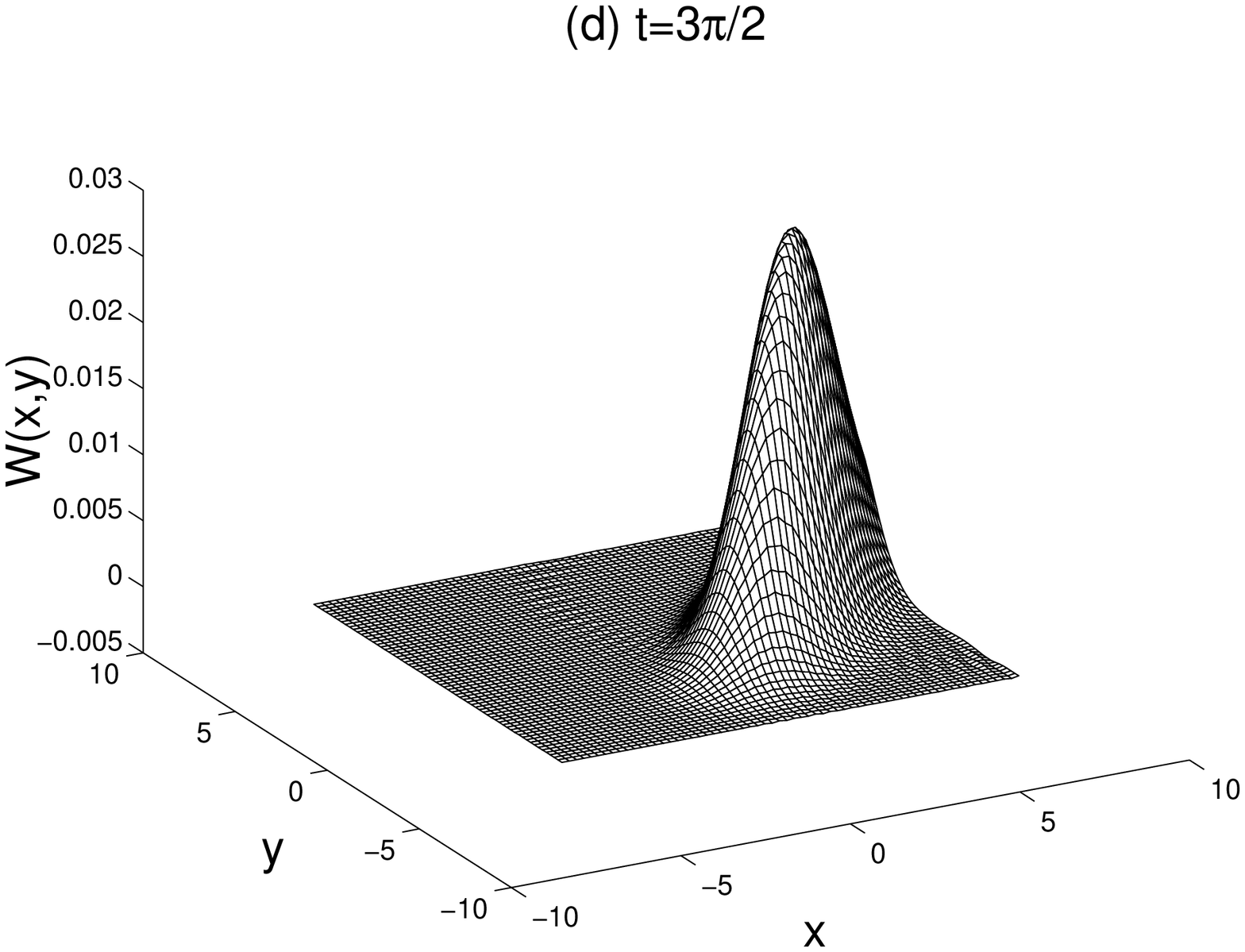}
\end{center}

\begin{center} 
\leavevmode 
\epsfxsize=8cm 
\epsfbox{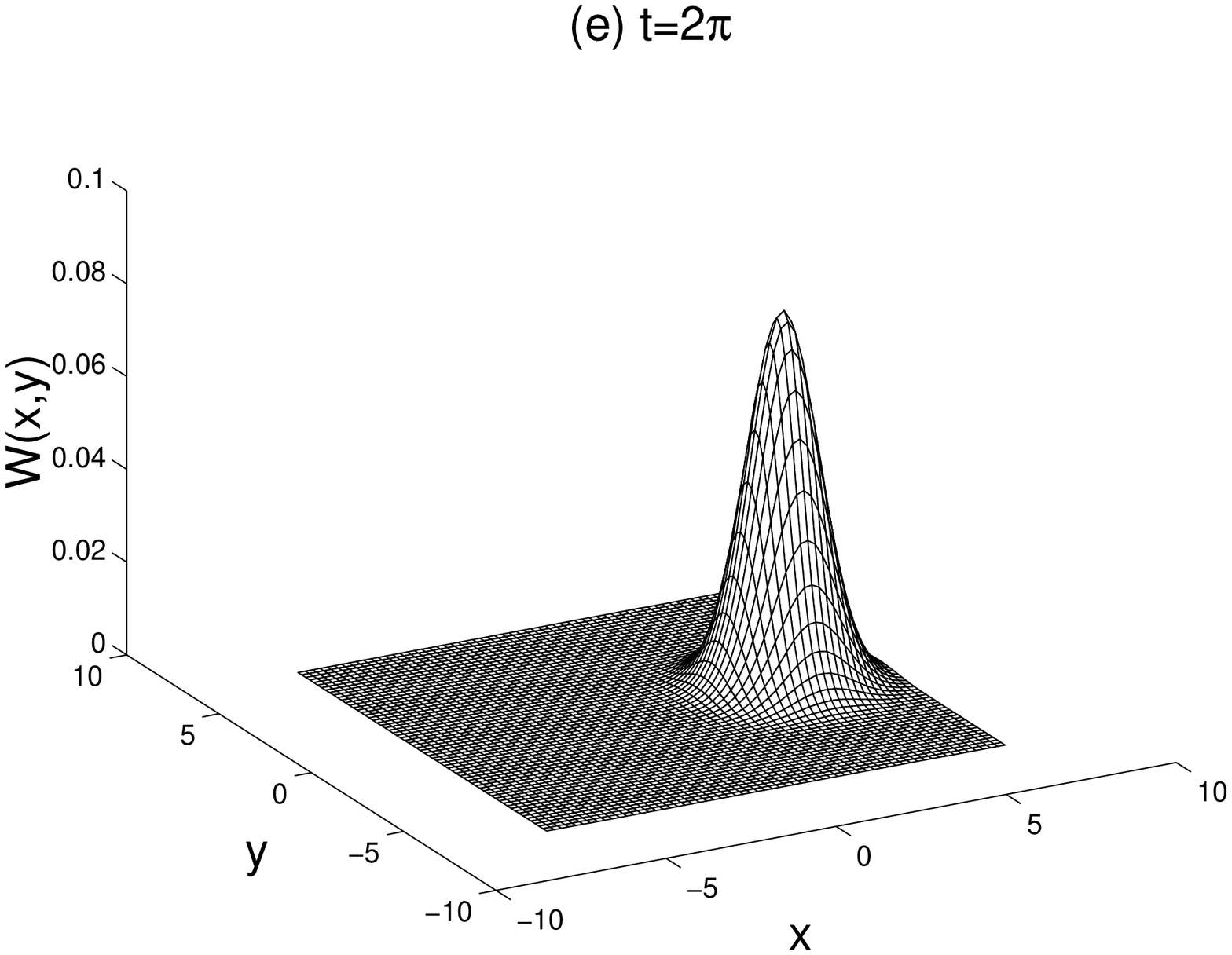}
\caption{\narrowtext The Wigner function, $W(x,y)$, of the mirror at various times $t$ of the system in the absence of damping. We have taken the initial coherent state amplitude of the cavity mode to be $\alpha=2$, and that of the mirror to be $\beta=2$. The scaled coupling parameter has been set at $k=0.5$. The quadratures $x$ and $y$ and the Wigner function $W(x,y)$ are given here in dimensionless form.}
\end{center}
 
\end{figure}

Thus, during undamped evolution, no nonclassical state of the mirror is generated. However, because of the entanglement of the mirror with the field, it is possible to generate nonclassical states of the
mirror by performing conditional measurements on the field. We discuss this in section~\ref{mcon}.

\section{Generation of nonclassical states of the cavity field}
We divide the discussion of the generation of nonclassical states of the cavity field  into three categories: (A) Multicomponent cats generated at $t=2\pi$ due to dynamics alone (i.e without any external intervention). Among these, a method of generating the 2-component Schr\"{o}dinger cat has been discussed in Ref.\cite{Tombesi}. (B) Entangled states of two or more light modes generated at $t=2\pi$ due to the dynamics alone. (C) Nonclassical states produced due to conditional measurements on the mirror. Of course the discussion here is by no means exhaustive and there remains the possibility of generating even more interesting states.

\subsection {Multicomponent cats}

At time $t=2\pi$ the state of the cavity field, as obtained from Eq.(5), is given by
\begin{equation}
  |\zeta\rangle_{\mbox{\scriptsize c}} = e^\frac{-|\alpha|^2}{2}\sum_{n=0}^\infty\frac{\alpha^n}{\sqrt{n!}}e^{i 2 \pi k^2 n^2 }|n\rangle_{\mbox{\scriptsize c}}
\end{equation}
Depending on the value of the parameter $k$, the state $|\zeta\rangle$ can be made equivalent to a variety of multicomponent cats. For $k=0.5$, 
\begin{eqnarray}
  |\zeta_2\rangle_{\mbox{\scriptsize c}} & = & e^\frac{-|\alpha|^2}{2}\sum_{n=0}^\infty\frac{\alpha^n}{\sqrt{n!}}e^{in^2 \frac{\pi}{2}}|n\rangle_{\mbox{\scriptsize c}}  \nonumber\\
    & = & e^\frac{-|\alpha|^2}{2} \left[\left(\frac{1+i}{2}\right) |+\alpha\rangle_{\mbox{\scriptsize c}} + \left(\frac{1-i}{2}\right)|-\alpha\rangle_{\mbox{\scriptsize c}} \right] \end{eqnarray}
which is a two-component Schr\"{o}dinger cat. For $k=(1/\sqrt{6})$ we get the three component cat,
\begin{equation}
  |\zeta_3\rangle_{\mbox{\scriptsize c}} = c_1 |-\alpha \rangle_{\mbox{\scriptsize c}} + c_2 |\alpha e^{i\frac{\pi}{3}} \rangle_{\mbox{\scriptsize c}} + c_3  |\alpha e^{-i\frac{\pi}{3}}\rangle_{\mbox{\scriptsize c}}
\end{equation}
where $c_1 =1$ and $c_2= -c_3 = (1+e^{i\pi/3})/(2i\sin(\pi/3))$ (not normalized). For $k=1/(2\sqrt 2)$,
\begin{equation}
  |\zeta_4\rangle_{\mbox{\scriptsize c}} = \frac{e^{i\frac{\pi}{4}}}{2}(|\alpha\rangle_{\mbox{\scriptsize c}} - |-\alpha \rangle_{\mbox{\scriptsize c}}) +  \frac{1}{2}(|i\alpha \rangle_{\mbox{\scriptsize c}} +|-i\alpha \rangle_{\mbox{\scriptsize c}})
\end{equation}
which is a four component cat (not normalized). Thus by adjusting the ratio of the coupling and the mirror frequency and thereby varying $k$ one can, in principle, obtain all these types of cats at time  $t=2\pi$. The Wigner functions of the cats produced by different values
of $k$ are shown below in Fig.(3). In this figure and all figures showing the Wigner function of the cavity field $x=a+a^\dagger$ and $y=-i(a-a^\dagger)$ are dimensionless quantities. Exactly the same type of states have been noted  before as arising in  the Kerr medium~\cite{Tanas,Gar,Ag}. It is indeed worth noting that creating such states in a cavity such as ours has intrinsic advantages because there is an extensive set of tomographic methods~\cite{tom1,tom2,tom3,tom4,tom5,tom6,tom7} that can be implemented to reconstruct the Wigner function in a cavity.
\begin{figure}

\begin{center} 
\leavevmode 
\epsfxsize=8cm 
\epsfbox{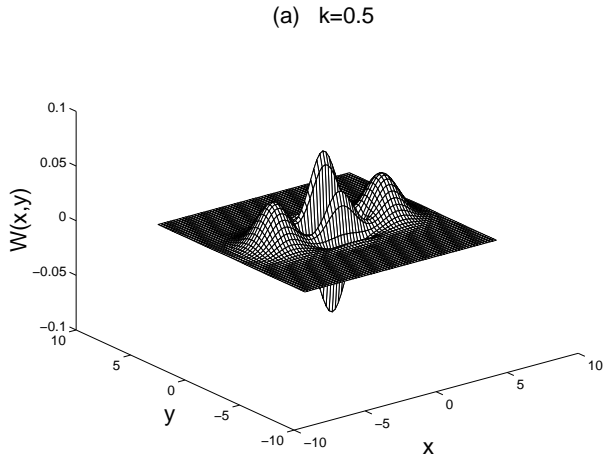}
\end{center} 
\begin{center} 
\leavevmode 
\epsfxsize=8cm 
\epsfbox{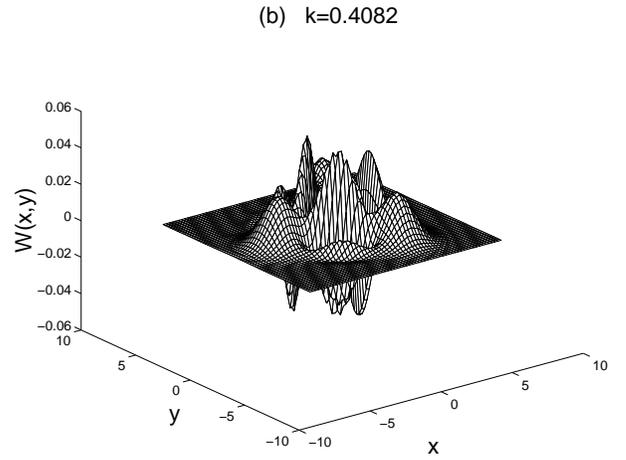}
\end{center}

\begin{center} 
\leavevmode 
\epsfxsize=8cm 
\epsfbox{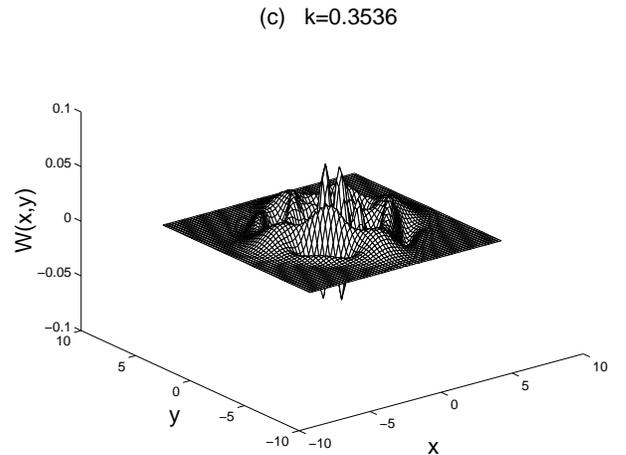}
\caption{\narrowtext The Wigner functions of the state of the cavity field at $t=2\pi$ are plotted here for various values of the scaled coupling parameter $k$. We have taken the initial coherent state amplitude of the cavity mode to be $\alpha=2$, and that of the mirror to be $\beta=2$. The quadratures $x$ and $y$ and the Wigner function $W(x,y)$ are given here in dimensionless form.}
\end{center}

\end{figure}

\subsection {Entangled states of two or more cavity modes}
\label{enta}
Two modes of light, each  separately interacting with the movable mirror, and with no direct coupling between them, can end up being in an entangled state at time $t=2\pi$ depending on the value of the parameter $k$. Let there be two different modes of light inside the cavity at $t=0$. For simplicity we assume them to have the same frequency and hence same value of the parameter $k$, but mutually orthogonal polarization directions. We assume them to be initially fully disentangled and prepared in coherent states. Let the initial state of the composite system of the mirror and the light be 
\begin{equation}
     |\Psi(0)\rangle = |\alpha_1\rangle_{\mbox{\scriptsize c1}} \otimes  |\alpha_2\rangle_{\mbox{\scriptsize c2}} \otimes |\beta \rangle_{\mbox{\scriptsize m}}               \end{equation}
where $|\alpha_1\rangle_{\mbox{\scriptsize c1}}$ and $|\alpha_2\rangle_{\mbox{\scriptsize c2}}$ are states of the first and second modes respectively, while $|\beta\rangle_{\mbox{\scriptsize m}}$ is the initial state of the mirror. The state of the two light modes evolves at $t=2\pi$ to the state (not normalized) 
\begin{equation}
  |\zeta_E \rangle =\sum_{n,m=0}^\infty\frac{\alpha^n_1\alpha^m_2 e^{i 2 \pi k^2 (n+m)^2 }}{\sqrt{n!}\sqrt{m!}}|n\rangle_{\mbox{\scriptsize c1}} \otimes |m\rangle_{\mbox{\scriptsize c2}} .
\end{equation}
 The state can be rewritten as 
\begin{equation}
  |\zeta_E \rangle =\sum_{m=0}^\infty\frac{\alpha_2^m}{\sqrt{m!}} e^{i k^2 m^2 2\pi} |m\rangle_{\mbox{\scriptsize c1}} \otimes |\xi(m,k)\rangle_{\mbox{\scriptsize c2}}
\end{equation}
where
\begin{equation}
  |\xi(m,k)\rangle_{\mbox{\scriptsize c2}} = \sum_{n=0}^\infty\frac{(\alpha_1 e^{i k^2 4\pi m})^n}{\sqrt{n!}}e^{i 2 \pi k^2 n^2 }|n\rangle_{\mbox{\scriptsize c2}}  .
\end{equation}
As is evident from the above equation, $|\xi(m,k)\rangle_{\mbox{\scriptsize c2}}$ are states of the same type as the state $|\zeta\rangle_{\mbox{\scriptsize c}}$ given by Eq.(7) with an amplitude $\alpha_1 e^{i k^2 4\pi m}$. Hence, for certain values of $k$ (such as 0.5 or $1/(\sqrt 6)$ or $1/(2\sqrt 2)$), number states of one mode become correlated with multicomponent cat states of the other mode. However, if $k$ is such that $(2 k^2  nm)$ is  an integer for all $n$ and $m$, then the state $|\xi(m,k)\rangle_{\mbox{\scriptsize c2}}$  becomes independent of $m$ and $|\zeta_E \rangle$ becomes a disentangled state.
Some simplification of Eq.(14) for k=0.5 results in the entangled state
\begin{equation}
 \begin{array}{rcrcl}
  |\zeta_E \rangle & = & (1+i) |+\alpha_1\rangle_{\mbox{\scriptsize c1}} \otimes |+\alpha_2\rangle_{\mbox{\scriptsize c2}} \nonumber \\
 & + & (1-i) |-\alpha_1\rangle_{\mbox{\scriptsize c1}}  \otimes  |-\alpha_2\rangle_{\mbox{\scriptsize c2}} .
 \end{array}
\end{equation}
 Note that production of entangled states in a Kerr medium, when the two modes of light have a direct interaction between them have been discussed before~\cite{Barn}; however this case is different in the sense that the modes become entangled because they both individually interact with the same movable mirror.\\
  There is no restriction on the number of modes that can be entangled using the above procedure (only each mode to be entangled has to be resonant with the cavity). It is interesting to demonstrate how eigenstates
of the operator $ (a_{\mbox{\scriptsize c1}} a_{\mbox{\scriptsize c2}}...a_{\mbox{\scriptsize cN}})^p $ where $ a_{\mbox{\scriptsize c1}}, a_{\mbox{\scriptsize c2}},..., a_{\mbox{\scriptsize cN}} $ are annihilation operators of  $ N $ different modes of light and $ p $ is an integer can be created in the cavity. As it is possible for resonant modes to have
frequencies that are integer multiples of each other, let us consider the case when there are $ N $ different modes of light in
the cavity, with their frequencies $ \omega_{\mbox{\scriptsize c1}} , \omega_{\mbox{\scriptsize c2}} , .... , \omega_{\mbox{\scriptsize cN}} $ being related by $ \omega_{\mbox{\scriptsize cj}} = \eta_j ~ \omega_{\mbox{\scriptsize c1}} $ where $ \eta_j $ are integers. Then from  Eq.(\ref{gpar}) it follows that the value of their $ k $ parameters will be related by $ k_j=\eta_j ~k_1 $. Hence if the initial state of the cavity modes and the mirror is 
\begin{equation}
     |\Psi(0)\rangle = |\alpha_1\rangle_{\mbox{\scriptsize c1}} \otimes  |\alpha_2\rangle_{\mbox{\scriptsize c2}} \otimes ....\otimes|\alpha_1\rangle_{\mbox{\scriptsize cN}} \otimes |\beta \rangle_{\mbox{\scriptsize m}}   ~          ,  \end{equation}
the final (at time $ t=2\pi $ ) composite state of the cavity modes will be
\end{multicols}
\vspace{-0.5cm}
\noindent\rule{0.5\textwidth}{0.4pt}\rule{0.4pt}{\baselineskip}
\widetext
\begin{equation}
\label{comp}
  |\zeta_{E_N} \rangle =\sum_{n_1,n_2,...,n_N=0}^\infty  e^{i 2 \pi {k_1}^2 ( \eta_1 ~n_1+\eta_2 ~n_2+...+\eta_N ~n_N)^2 } ~\prod_{j=0}^N \frac{\alpha^{n_j}_j }{\sqrt{{n_j}!}} ~ |n_j\rangle_{\mbox{\scriptsize cj}} .
\end{equation}
As the term $ e^{i 2 \pi {k_1}^2 ( \eta_1 ~n_1+\eta_2 ~n_2+...+\eta_N ~n_N)^2 } $ cannot be split up into a product of the form $ f_1(n_1) f_2(n_2) ... f_N(n_N) $, for arbitrary values of $ k_1 $, $ |\zeta_{E_N} \rangle $ can, in general, be an entangled state depending on the value of $ k_1 $. Now consider the case when $  2 \pi {k_1}^2 = \pi /p $. We have
\begin{equation}
 (a_{\mbox{\scriptsize c1}} a_{\mbox{\scriptsize c2}}...a_{\mbox{\scriptsize cN}})^p |\zeta_{E_N} \rangle = \sum_{n_1,n_2,...,n_N=p}^\infty  e^{i 2 \pi {k_1}^2 ( \eta_1 ~n_1+\eta_2 ~n_2+...+\eta_N ~n_N)^2 } ~\prod_{j=0}^N \frac{\alpha^{n_j}_j }{\sqrt{{n_j-p}!}} ~ |n_j-p\rangle_{\mbox{\scriptsize cj}}~. 
\end{equation}
Redefining $ n_j $ in the above equation as $ n_j + p $ , recasting the equation
in terms of the new $ n_j $ and using the fact that $ \eta_j $ are integers, one
gets   
\begin{equation}
 (a_{\mbox{\scriptsize c1}} a_{\mbox{\scriptsize c2}}...a_{\mbox{\scriptsize cN}})^p |\zeta_{E_N} \rangle = (\alpha_1 \alpha_2 ... \alpha_N)^p e^{i \pi p (\eta_1+\eta_2+...+\eta_N)^2} ~|\zeta_{E_N} \rangle ~.
\end{equation}
Hence $ |\zeta_{E_N} \rangle $ are eigenstates of  the operator $ (a_{\mbox{\scriptsize c1}} a_{\mbox{\scriptsize c2}}...a_{\mbox{\scriptsize cN}})^p $, generalizations of the well known even and odd coherent states which are eigenstates of squares of annihilation operators~\cite{Buzek}.

\begin{multicols}{2}

\subsection {Nonclassical states produced by conditional measurements on the mirror}
\label{fock}
Because of the entanglement of the mirror state with the state of the cavity field during evolution, any measurement on the mirror will project the cavity field to some state which we can determine. We consider here measurements of the mirror's position, $x=b+b^\dagger$ (i.e, $x$ represents actual position of the
mirror multiplied by $(\sqrt{2m\omega_m/\hbar})$, $m$ being the mass of the mirror), at time $t=\pi$, as the entanglement is most pronounced at that time. Such a measurement will project the state of the light to
\begin{equation}
\label{fsup}
  |\eta(x)\rangle_{\mbox{\scriptsize c}} = N\sum_{n=0}^\infty\frac{\alpha^n}{\sqrt{n!}}e^{i \pi k^2 n^2 } \langle x|\phi_n (\pi)\rangle
|n\rangle_{\mbox{\scriptsize c}} .
\end{equation}
Thus for different values of the mirror position $x$, different states of the cavity field are produced. Let us choose parameters  $k=1, \alpha=2$ and $\beta=2$. Then the peak of the gaussian function $\langle x|\phi_n (\pi)\rangle$ lies at $x_n=4n-4$ and its width is $\delta = 1$. Thus the distance 
between the peaks of  $\langle x|\phi_n (\pi)\rangle$ and  $\langle x|\phi_{n+1} (\pi)\rangle$, that is, $ x_n-x_{n+1}=4 $ , is greater than the width $ \delta $.Hence, if the value obtained for $x$ obtained as a result of the measurement is near $x_n$, the most dominant constituent of the state $|\eta(x)\rangle_{\mbox{\scriptsize c}}$ is $|n\rangle_{\mbox{\scriptsize c}}$. In this way, we may expect to produce nonclassical states which are very close to number
states. For example, if the value obtained for $x$ is near 0 (which is near $x_1$) the cavity field is projected to the state (unnormalized)
\begin{equation} 
  |\eta(0)\rangle \approx e^{-4}|0\rangle_{\mbox{\scriptsize c}} - 2|1\rangle_{\mbox{\scriptsize c}} + 2\sqrt{2}e^{-4}|2\rangle_{\mbox{\scriptsize c}} + O(10^{-7})
\end{equation}
which is shown in Fig.(4a). Similarly when $x$ is near $x_5 \approx 16$, we get a state very close to the number state $|5\rangle$ as shown in Fig.(4b). When $x$ is inbetween $x_n$ and $x_{n+1}$ we get essentially a superposition of states $|n\rangle$ and $|n+1 \rangle$ as illustrated in Fig.(4c) for $n=2$ (the state in the figure is approximately $-2 |1\rangle_{\mbox{\scriptsize c}} + 2\sqrt{2} |2\rangle_{\mbox{\scriptsize c}}$, not normalized). Fock state superpositions and their nonclassical properties have been discussed in~\cite{Wod}.
\begin{figure} 

\begin{center} 
\leavevmode 
\epsfxsize=8cm 
\epsfbox{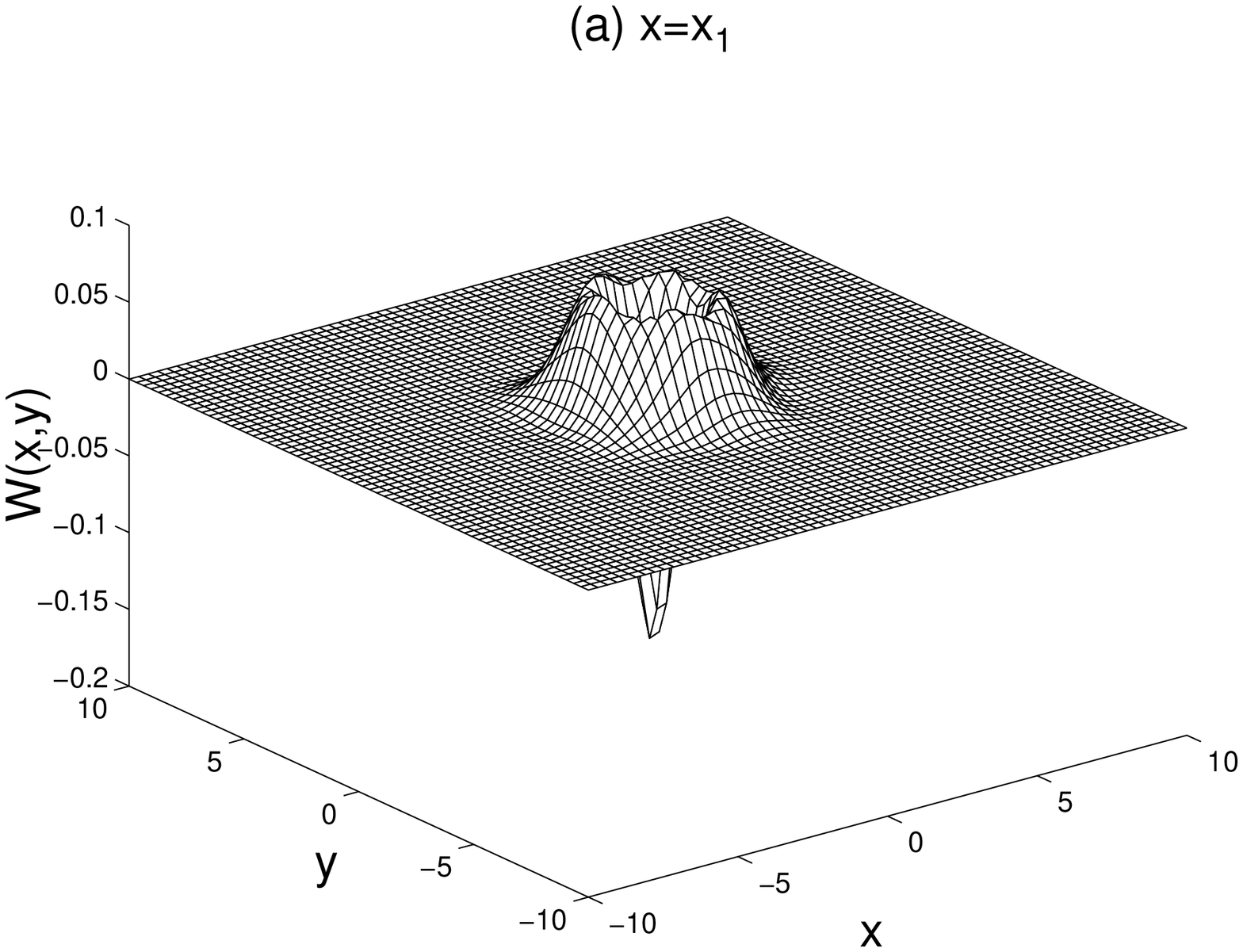}
\end{center} 
\begin{center} 
\leavevmode 
\epsfxsize=8cm 
\epsfbox{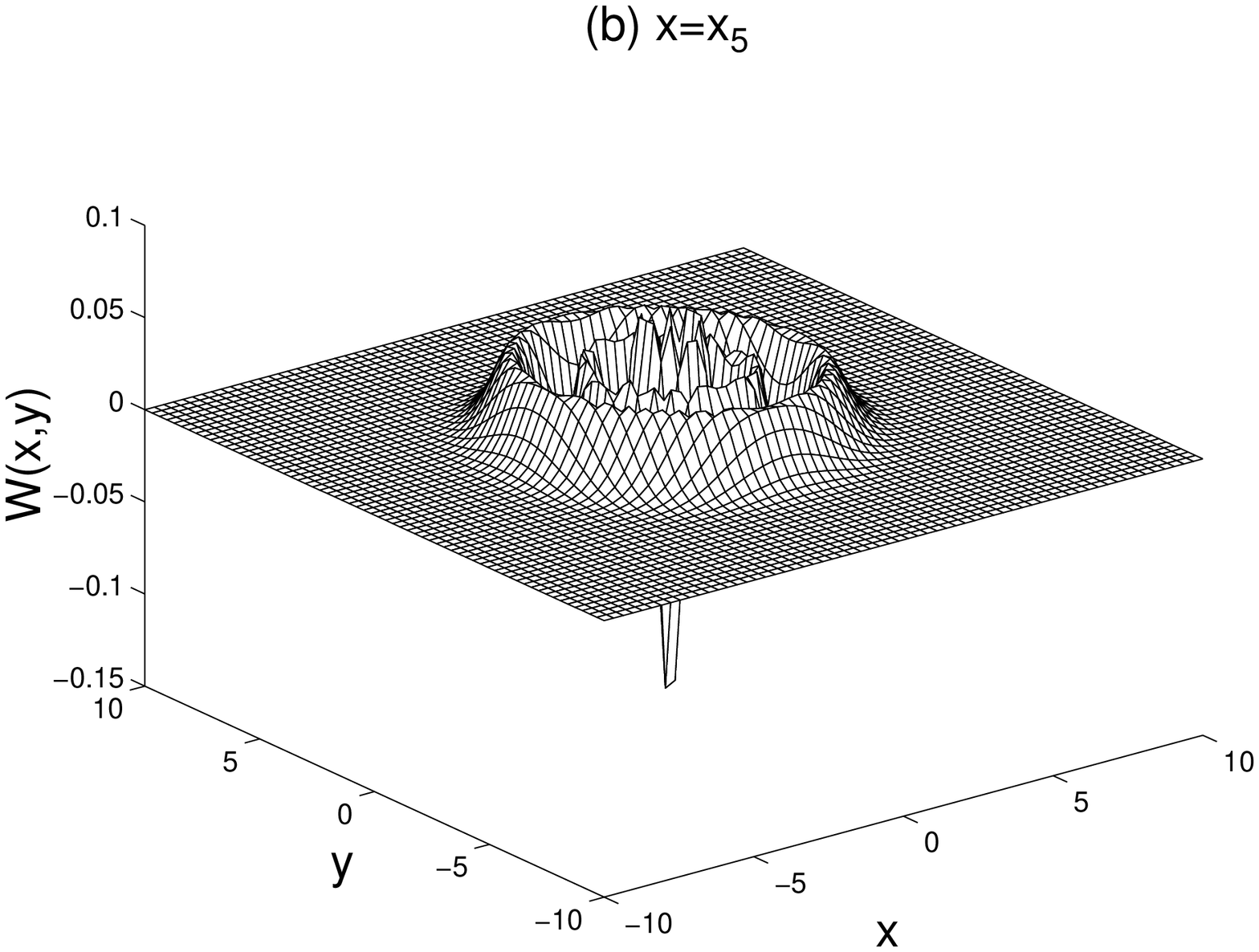}
\end{center}

\begin{center} 
\leavevmode 
\epsfxsize=8cm 
\epsfbox{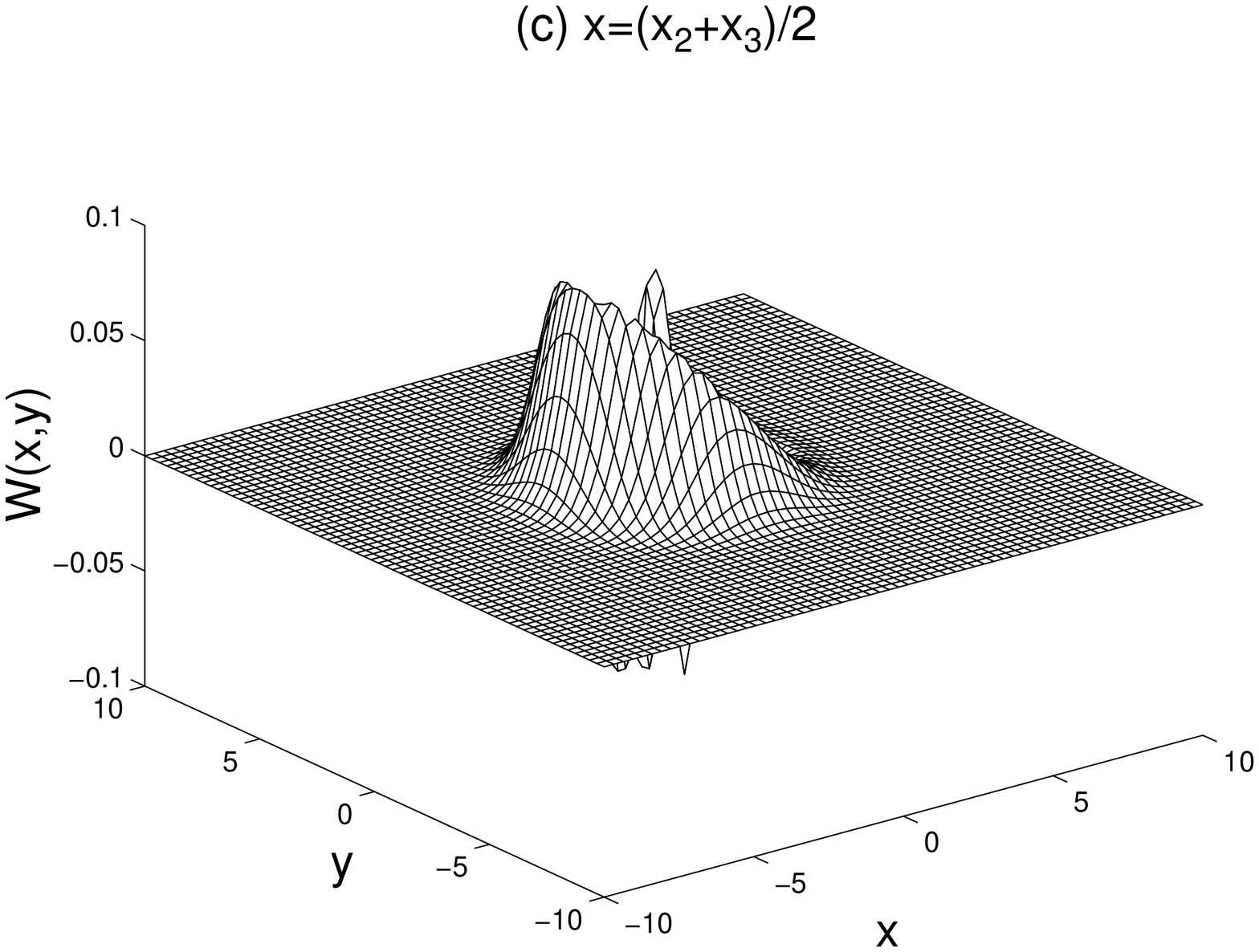}
\caption{\narrowtext The Wigner functions of the state to which the cavity field is projected for
various results, $x$, of a measurement of the position of the mirror. We have taken the initial coherent state amplitude of the cavity mode to be $\alpha=2$, and that of the mirror to be $\beta=2$. The scaled coupling parameter is $k=1$, and the measurment is made at time $t=\pi$. The quadratures $x$ and $y$ and the Wigner function $W(x,y)$ are given here in dimensionless form.}
\end{center}

\end{figure}
It is evident from the above discussion that the generation of the above type of states solely rely on the narrowness of the coherent state width in comparison to the spatial separation between the peaks of
the gaussians  $\langle x|\phi_n (\pi)\rangle$ and  $\langle x|\phi_{n+1} (\pi)\rangle$. As this spatial separation can always be increased by increasing
the value of the parameter $k$, the proximity of the states produced to Fock
states and their superpositions can always be improved. 

    We note that the first two types of states, that is, the cats and the entangled states, rely on the Kerr type term in the time evolution and hence they can be generated in a Kerr medium as well. In particular, generation of cat states in a Kerr medium was suggested some time ago~\cite{Tanas,Gar}. However, the generation of the third type of states (i.e the number states and their superpositions) do not appear to have an analogue in the Kerr medium because they are entirely dependent on the entanglement of the states of the mirror and the cavity field during evolution.

\section{Generation of nonclassical states of the mirror from conditional measurements} \label{mcon}
The system also offers the opportunity to create quite different nonclassical states of the mirror by conditional measurements on the $x$ quadrature of the light field. As the spatial separation between the coherent state components
of the mirror is maximum at $ t=\pi $,  measurements at this instant are likely
to produce the most nonclassical states of the mirror. A measurement of the $x$ quadrature of the light field at this instant of time  projects the mirror state to
\begin{equation}
\label{mir}
  |\Phi(x)\rangle_{\mbox{\scriptsize m}} = e^\frac{-|\alpha|^2}{2}\sum_{n=0}^\infty\frac{\alpha^n}{\sqrt{n!}}e^{ik^2 n^2 \pi}\langle x|n\rangle |\phi_n (\pi)\rangle_{\mbox{\scriptsize m}} .
\end{equation}
As the $|\phi_n (\pi)\rangle_{\mbox{\scriptsize m}}$ are coherent states with a different amplitude for each value of $n$, $|\Phi(x)\rangle_{\mbox{\scriptsize m}}$ is a {\em superposition of spatially separated coherent states} and as such an entirely nonclassical state of the mirror. Moreover, as the parameter $k$ is increased, the separation between the coherent components $ |\phi_n (\pi)\rangle_{\mbox{\scriptsize m}} $ increases. So by varying $ k $, one can control the macroscopic distinguishability of the states involved in the superposition. In Fig.(5) we have plotted the Wigner function for the state of the mirror produced when  $k=1$ and the measurement result is $x=0$.  This state is distinctly nonclassical and looks somewhat like a Schr\"{o}dinger cat state with unequal amounts of each coherent component.
\begin{figure}
\begin{center} 
\leavevmode 
\epsfxsize=8cm 
\epsfbox{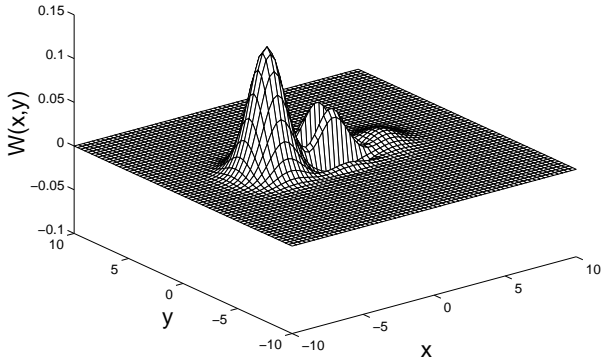}
\caption{\narrowtext The Wigner function of the state to which the mirror is projected when a measurement of the $x$ quadrature of the cavity field at time $t=\pi$ gives a value 0. We have taken the initial coherent state amplitude of the cavity mode to be $\alpha=0.8$, and that of the mirror to be $\beta=2$.The scaled coupling parameter is $k=1$. The quadratures $x$ and $y$ and the Wigner function $W(x,y)$ are given here in dimensionless form.}
\end{center}
\end{figure}

The importance of the above procedure stems from the observation that there is already a considerable amount of literature, both theoretical~\cite{gou} and experimental~\cite{Monroe,wein}, focussed on methods to prepare atoms in nonclassical motional states. So it is only natural to expect that placing a more massive object in such states should be an issue of serious consideration. The ability to place macroscopic objects in non-classical states may even posses applications. Hollenhorst, for example, has shown that placing gravitational wave detectors in squeezed states leads to higher sensitivity~\cite{Hollenhorst}. Allthough schemes relying on momentum transer to a massive object from several microscopic objects (such as atoms or neutrons) have been suggested~\cite{self}, these are definitely difficult to implement experimentally. Thus the scheme described in this section offers a possible direction from which the production of nonclassical states of a macroscopic object may be approached.  Obviously, these states will decohere very rapidly, and their realization may enable some tests of decoherence models~\cite{dec}. The three main issues which will need to be addressed to bring this scheme to a practical level are: (1) How well isolated the movable mirror can be made , (2) how to perform tomography of the states produced and (3) how to perform instantaneous measurements of the $x$ quadrature of the field inside the cavity. We leave the treatment of these issues for the future.

\section{The effect of environment induced decoherence of the mirror on the states of the cavity field} \label{damp}
In Ref.~\cite{Tombesi}, the decoherence of the cavity field was treated as entirely due to the leaking of light through the cavity mirrors. Here, we investigate the complementary case i.e. when the photon leakage from the cavity is almost absent, but damping of the mirror's motion is significant. It is somewhat artificial to assume no leakage of light from the cavity, but {\em in principle} the rate of mirror damping can be made a few orders of magnitude more than the rate of damping due to the leakage of light by choosing mirrors of sufficiently high reflectivity. We do not try to address issues of practical feasibility in this paper, however.

As the state of the cavity field is entangled with the state of the mirror at times between $0$ and $2\pi$, it is expected that the decoherence of the mirror will induce a decoherence of the cavity field. However, its effect on the state of the cavity field is expected to be {\em nontrivial}.  During the standard decoherence due to leaking of light from the cavity, the coherent states of light form the relevant pointer basis~\cite{point}. However, in the case considered here, coherent states of the mirror are in one to one correspondence with number states of the cavity field. So if decoherence of the mirror forces it towards a coherent state basis, it will induce a decoherence of the cavity field towards the number state basis. This decoherence will leave some imprint on the state produced in the cavity at time $t=2\pi$.

The equation which governs the decoherence of the mirror depends on the way the mirror is coupled to its environment. Here, for simplicity, we assume that the mirror is amplitude coupled to the environment at zero temperature, which implies the master equation~\cite{Walls1}
\begin{equation}
\label{master}
 \frac { d\rho (t)}{dt} = -\frac{i}{\hbar} [H,\rho(t)] +\frac{\gamma}{2} (2 b \rho (t) b^\dagger -  b^\dagger b  \rho (t) - \rho (t) b^\dagger b)
\end{equation}
where $H$ is the Hamiltonian of Eq.(\ref{g}). Since we are using a scaled time $t$, $\gamma$ in the above equation is the usual damping constant (that is, the reciprocal of the dissipation timescale) divided by the mirror frequency $\omega_m$. Eq.(\ref{master}) has the feature that it singles out coherent states as the pointer basis~\cite{point}. It is expected that natural decoherence of macroscopic objects should force them towards a coherent state basis (to conform with classical reality), and  Eq.(\ref{master}) accomplishes exactly this. So we expect that the solution of Eq.\ref{master} will atleast give the basic features of the effect of mirror decoherence on the cavity states correctly. A similar justification has been given for the use of the above type of decoherence in modelling quantum measurement in Ref.\cite{Walls2}. A more appropriate model for the decoherence of the mirror is quantum Brownian motion~\cite{dec}, but it is also more difficult to solve for our system of interest. Work is in progress to solve the quantum Brownian motion for our system numerically, while here we present the analytic solution for the evolution of our system when decoherence is described according to Eq.(\ref{master}). This solution is expected to have most features (at least the localization of the cavity field towards a number state basis) similar to the solution in the case of quantum Brownian motion because both drive the mirror state to a mixture of coherent states.

The technique we use to solve Eq.(\ref{master}) is to apply the unitary evolution and the nonunitary (decohering) evolution alternately for short instants of time $\Delta t$, and then take the limit as $\Delta t \rightarrow 0$. For simplicity, we assume the initial coherent state of the mirror to be the vacuum state. The solution for this initial condition is evaluated in Appendix B, and the result is
\end{multicols}
\noindent\rule{0.5\textwidth}{0.4pt}\rule{0.4pt}{\baselineskip}
\widetext
\begin{equation}
\label{compl1}
  \rho (t) = e^{-|\alpha|^2}\sum_{n=0,m=0}^\infty\frac{\alpha^n{\alpha^*}^m}{\sqrt{n! m!}}e^{ik^2 (n^2-m^2) (t-\sin{t})} e^{-D(n,m,\gamma, t)} |n\rangle \langle m |\otimes |\phi_n (\gamma,t)\rangle \langle \phi_m (\gamma, t) | 
\end{equation}
where  
\begin{equation}
\label{compl2} 
  \phi_n (\gamma,t) = \frac{i k n}{i+\gamma/2 } (1- e^{-(i+\gamma/2)t}) ,
\end{equation}
are the amplitudes of the coherent states of the mirror and 
\begin{equation}
\label{compl3}
 D(n,m,\gamma, t) = \frac{k^2 (n-m)^2 \gamma}{2 (1+\gamma^2 /4) } \big[ t + \frac{1-e^{-\gamma t}}{\gamma} - (\frac{(e^{(i-\gamma/2)t}-1)}{i-\gamma/2}
-\frac{(e^{-(i+\gamma/2)t}-1)}{i+\gamma/2}) ] .
\end{equation}
\begin{multicols}{2}
The term $ e^{-D(n,m,\gamma, t)} $ in Eq.(\ref{compl1}) is responsible for
decoherence.Note that as $\gamma \rightarrow 0$, we have $D(n,m,\gamma, t) \rightarrow 0$ and $\phi_n (\gamma,t) \rightarrow \phi_n (t) $  where  $\phi_n (t)$ is that given by Eq.(\ref{phi1}), with $\beta$ set equal to zero. In other words, the solution given by Eq.(\ref{compl1}) reduces to the undamped solution given by Eq.(\ref{ro1}).
The fact that the mirror is in a mixed state and does not return to a pure state at $t=2\pi$ as in the undamped case is illustrated in Fig.(6) in a plot of the linear entropy of the mirror's state versus time.
\begin{figure} 
\begin{center} 
\leavevmode 
\epsfxsize=8cm 
\epsfbox{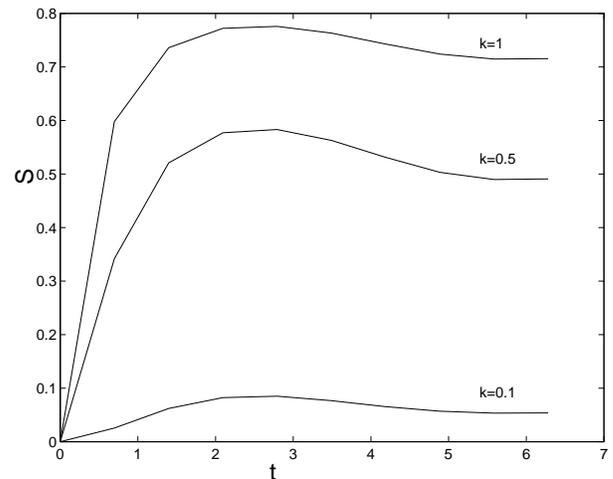}
\caption{\narrowtext The linear entropy, $S$, of the mirror state in the presence of damping. This is plotted here as a function of time, and for various values of the scaled coupling parameter $k$. We have taken the initial coherent state amplitude of the cavity mode to be $\alpha=2$, and that of the mirror to be $\beta=0$. We have chosen the damping constant to be $\gamma=1$. The entropy fails to return to zero because of the entanglement of the system with the environment. Both the scaled time $t$ and the linear entropy $S$ are dimensionless quantities.}
\end{center}
\end{figure}

The Wigner function for the light field at $t=2\pi$ for various values of our scaled damping constant $\gamma$ and $k=0.5$ (that is, when a Schr\"{o}dinger cat is expected in the absence of any decoherence) is given in Fig.(7). Two features concerning the type of decoherence considered here become clear from these figures:
\begin{figure}

\begin{center} 
\leavevmode 
\epsfxsize=8cm 
\epsfbox{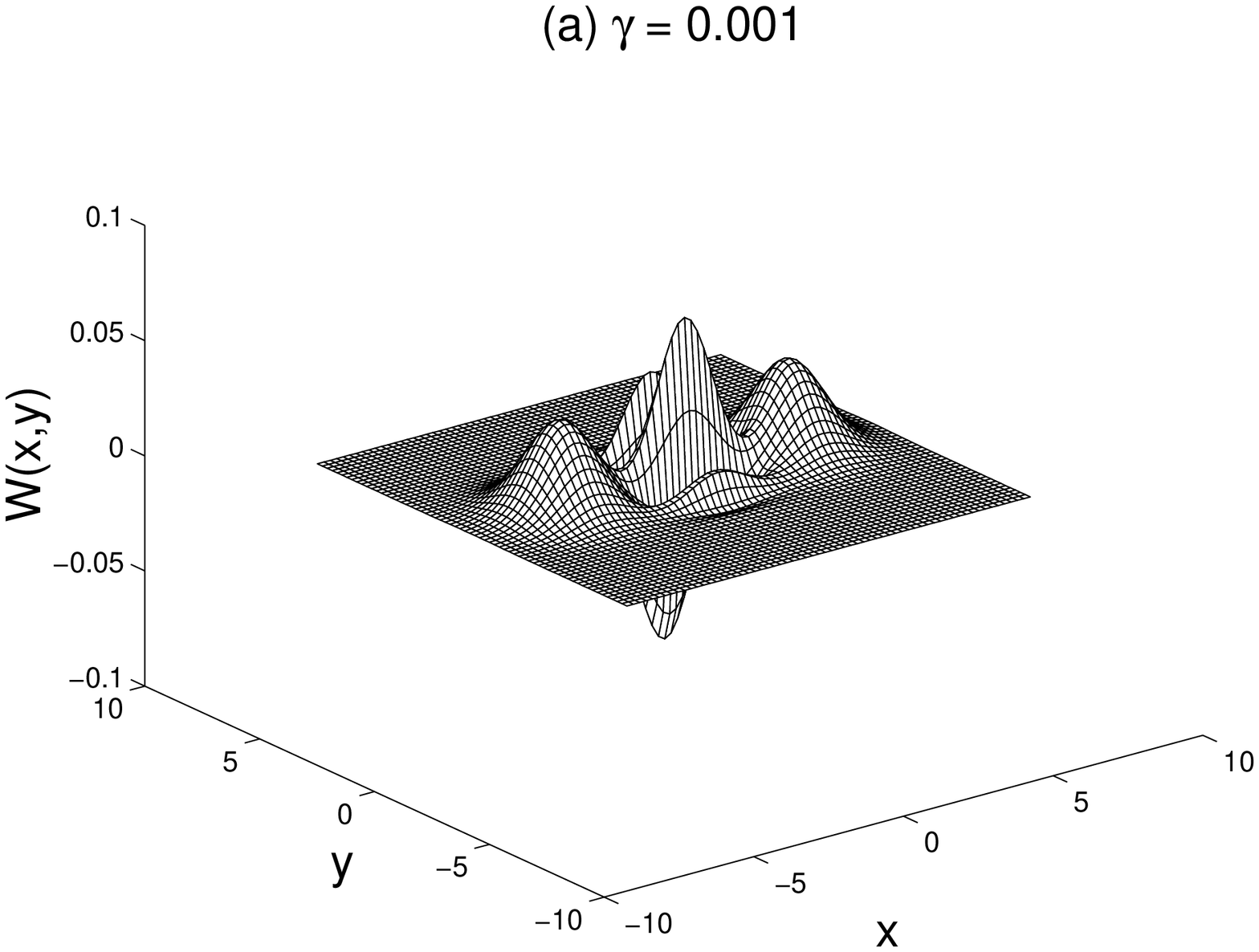}
\end{center} 
\begin{center} 
\leavevmode 
\epsfxsize=8cm 
\epsfbox{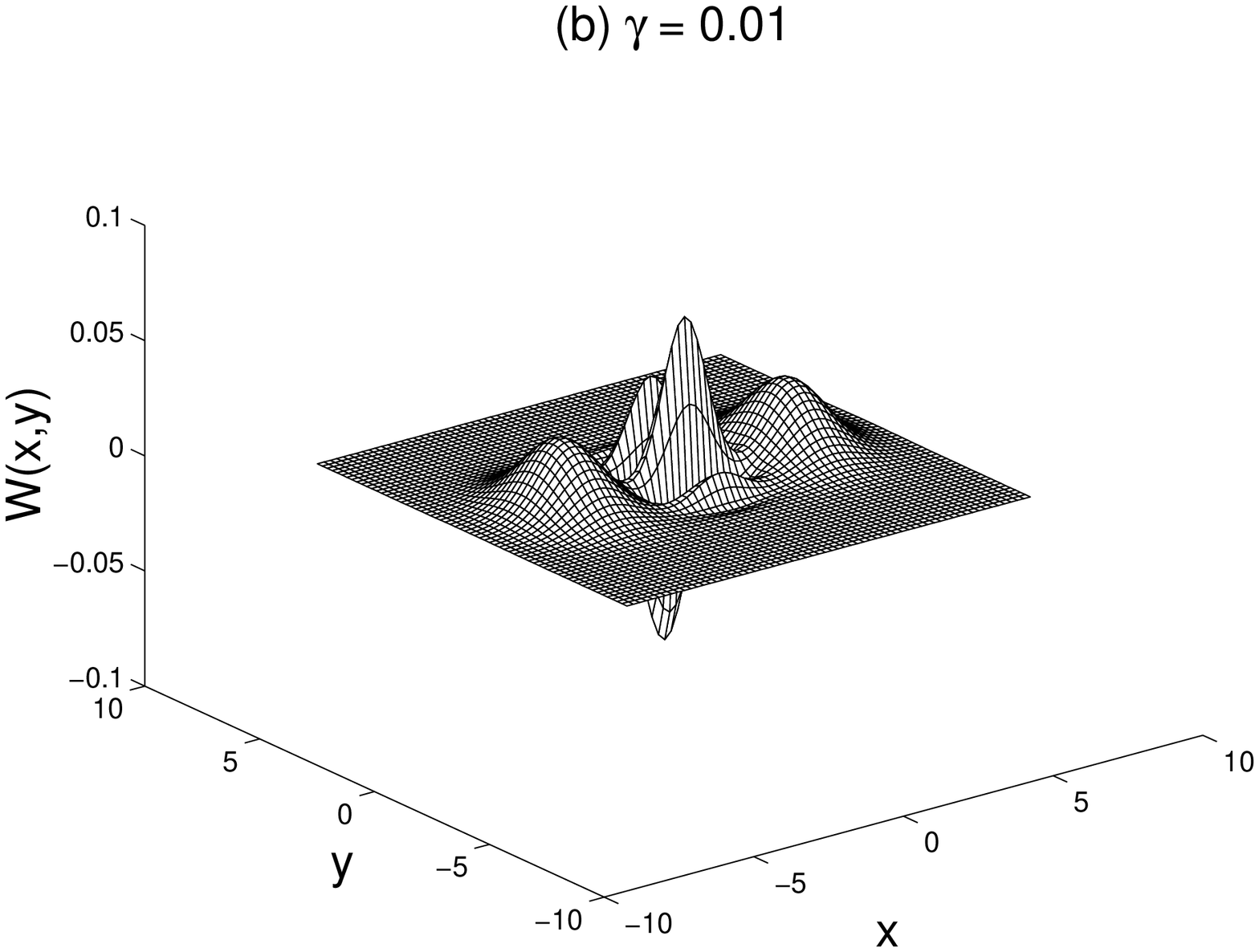}
\end{center}
\begin{center} 
\leavevmode 
\epsfxsize=8cm 
\epsfbox{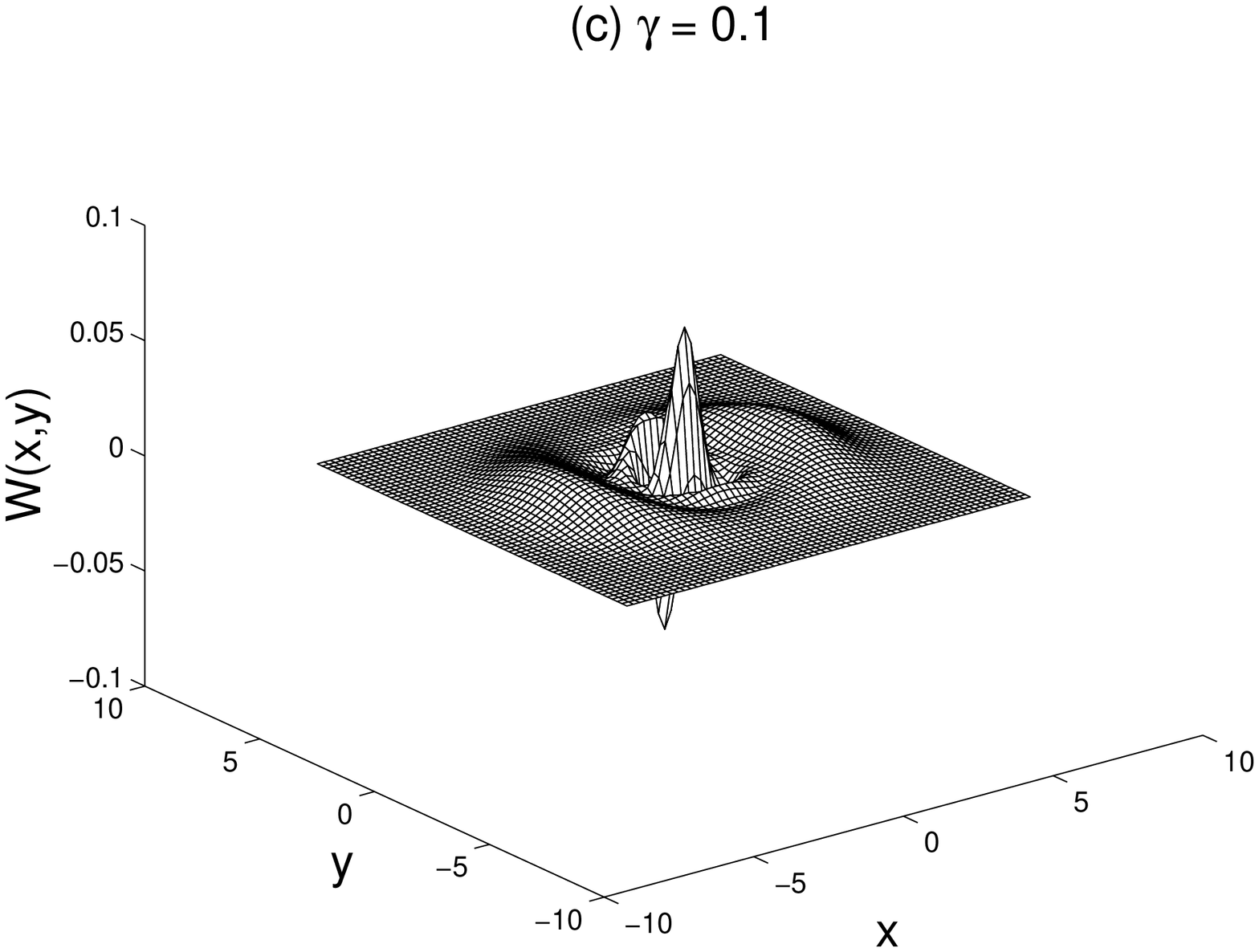}
\end{center} 
\begin{center} 
\leavevmode 
\epsfxsize=8cm 
\epsfbox{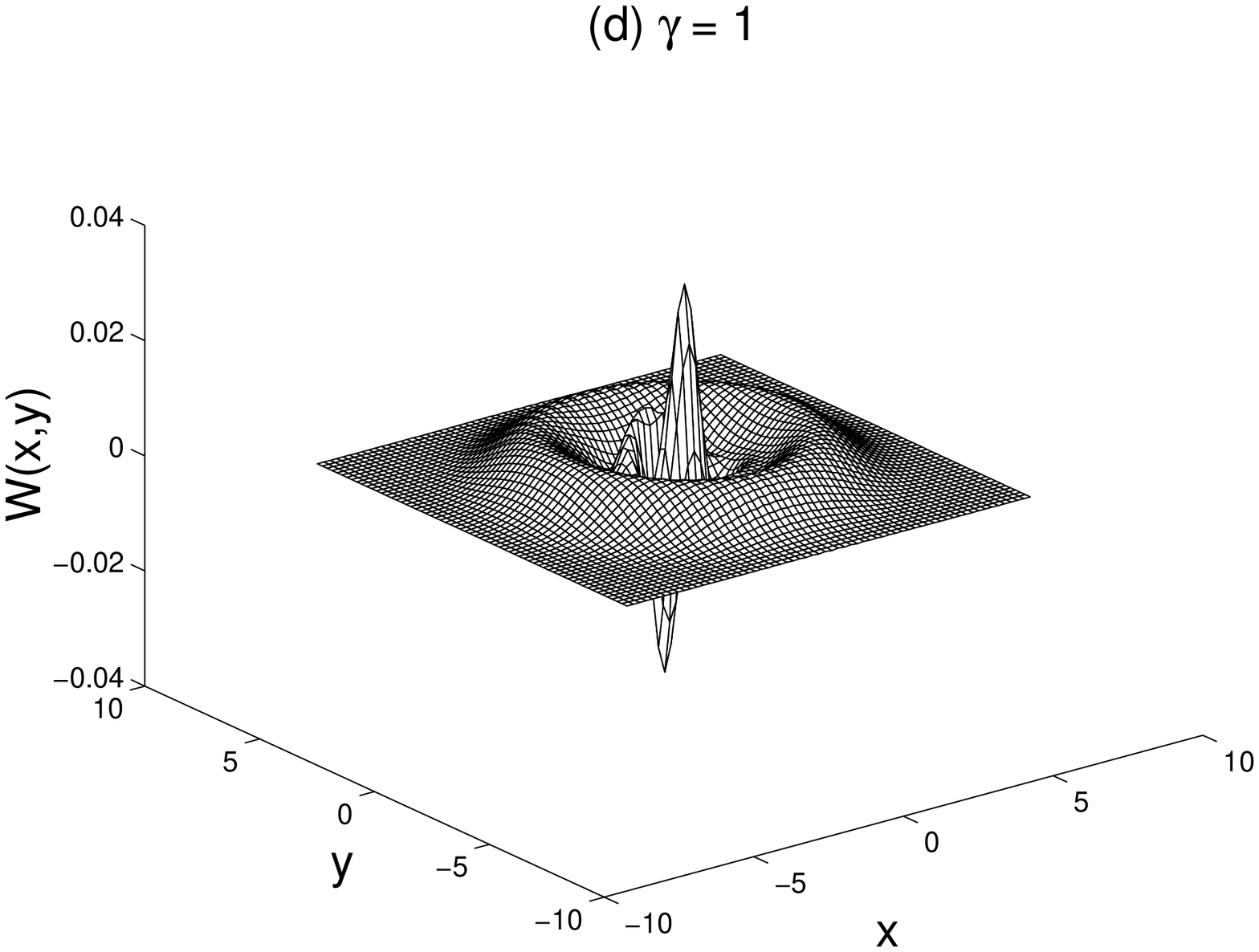}
\caption{\narrowtext The Wigner function $W(x,y)$ of the cavity field at time $t=2\pi$ for various values of the motional damping constant, $\gamma$, of the mirror. We have taken the initial coherent state amplitude of the cavity mode to be $\alpha=2$, and that of the mirror to be $\beta=2$. The scaled coupling parameter is $k=0.5$. The quadratures $x$ and $y$ and the Wigner function $W(x,y)$ are given here in dimensionless form.}
\end{center}

\end{figure}

(1) Note that from Fig.7(a), it is evident that even for a value of scaled  $\gamma$ as high as 0.001 (that is, unscaled $\gamma$ of 1), we have almost no decoherence of the Schr\"{o}dinger cat at all. Thus, unless the mirror is quite heavily damped it has almost no effect on the states produced inside the cavity. A possible cause for this is that the efficiency of decoherence depends on the separation between the coherent states of the mirror corresponding to different number states of the field. This separation is not constant, but oscillates between zero and a maximum value of $2k(n-m)$. The decoherence process, does not, therefore, get the chance to act with as much efficiency as it would have if the mirror was not a harmonic oscillator. Speaking more mathematically, Eq.(\ref{compl3}) implies that $ D(n,m,\gamma, t) $ is proportional to
$ k^2 \gamma $ and not just $\gamma $. As increasing the frequency decreases
$k$, it also decreases the rate of decoherence even if $\gamma$ (which relates
to the isolation of the system) is held constant. In terms of the absolute time  $ \tau $ (obtained by dividing the scaled time $t$, which we had been using, by $\omega_m$), the time scale $ \tau_d $ of decoherence  depends on the
frequency as (from Eqs.(\ref{gpar}), (\ref{compl3}) and definition of $k$)
\begin{equation}
 \tau_d \propto  (\omega_m)^3
\end{equation}
Thus controlling $ \omega_m $ offers an alternate way to control the decoherence timescale of our system.

   It is worthwhile to mention that an analogous situation has been pointed out in Ref.\cite{kni} in the context of the Jaynes Cummings model, where atomic spontaneous emmission has a much weaker effect on revivals of atomic inversion than the cavity field damping. In that case, however, the explanation was  quite different, namely, that  spontaneous emmission is  independent of the cavity field intensity, while cavity damping is intensity dependent. This logic will also apply to our system when the intensity of the cavity field is large. The term which causes decoherence, $D(n,m,\gamma)$, is completely independent of the intensity of the cavity field. So as this intensity increases, the photon damping begins to dominate over the mirror's motional damping, as far as influencing the final state in the cavity is concerned.

  (2) The interference peak of the Schr\"{o}dinger cat, which is a primary feature of this cat, is not significantly lowered by the decoherence process (See Figs.7(b)-7(d)), allthough there is a phase diffusion of the state. The reasoning for this is the simple fact that the decoherence is towards the number state basis, which preserves the photon number distribution. The photon number distribution at $t=2\pi$ is same for both the decohered state given by Eq.(\ref{compl1}) and the undecohered state given by Eq.(\ref{ro1}). This means that one of the primary signatures of the Schr\"{o}dinger cat, namely, an oscillating photon number distribution (that is, the probability of odd photon numbers being zero for an even coherent state and vice versa) is the same for the final state produced at $t=2\pi$, irrespective of whether there was decoherence or not. So the decoherence considered here is not acting to eliminate all nonclassical properties of the field that would be produced without decoherence. It does destroy the phase information, but maintains the nonclassicality of the number distribution. This is precisely the reason why the interference peak of the Schr\"{o}dinger cat does not appear to be destroyed even when scaled $\gamma$ is as high as unity, as depicted in Fig.7(d).
 
So far we have only considered how the generation of the Schr\"{o}dinger cat is affected in the presence of the mirror's motional damping. Let us now
briefly pause to consider how the generation of the other nonclassical states of the cavity field may be affected. The entangled states of two or more cavity modes mentioned in Section.\ref{enta} are generated by a procedure identical
to the generation of the multicomponent cats and hence should have similar
patterns of decoherence. On the other hand, the Fock states mentioned in 
Section.\ref{fock} will not atall be affected by the type of decoherence considered here, as it drives the cavity field towards a mixture of Fock states.
Moreover, even after their production they will be very stable because, neither
the Hamiltonian evolution, nor the decohering evolution destroys a Fock state
of the cavity field. Ofcourse one will have to remember that in a realistic
case the photon leakage will also be present, which does destroy Fock states \cite{Tanas,Gar}. Lastly, comes the superpositions of two successive Fock states, also
described in Section.\ref{fock}. These will, ofcourse, be seriously affected
as the decoherence process will be destroying the coherence between successive
Fock states. As the generation of the nonclassical states of the mirror depend crucially on the coherence between the  different $ |n\rangle_{\mbox{\scriptsize c}} \otimes |\phi_n (t)\rangle_{\mbox{\scriptsize m}} $ components of the system's state, it will also be affected. In fact, at the very instant of
generation of a nonclassical state of the mirror of the type given by Eq.(\ref{mir}), the coherence between any two of its
components like $ |\phi_n (\pi)\rangle_{\mbox{\scriptsize m}} $ and  $ |\phi_m (\pi)\rangle_{\mbox{\scriptsize m}} $,   will be  $ e^{-D(n,m,\gamma, \pi)}  $ 
of the undamped value.

    We have already mentioned that the decoherence timescale of the system may
be increased by increasing the frequency of the mirror. But whether that
actually helps in the generation of any of the nonclassical states described
in this paper requires further scrutiny. For example, the multicomponent
cats are generated only when $ k^2 t $ reaches a certain value. But  $ D(n,m,\gamma, t) $ depends precisely on $ k^2 t $ and not simply on time. Thus increasing the frequency will ofcourse increase the decoherence timescale, but also increase the timescale for the production of the cat states proportionately, so that they are decohered by the same amount whenever they are produced. For the nonclassical states of the mirror described by Eq.(\ref{mir}),
the situation is somewhat better, but not without its problems. By decreasing $k $ (through $\omega_m $), one not only decreases the rate of decoherence but also decreases the spatial separation between the coherent state components involved in the superposition. Hence the components in the superposition may become more coherent, but they also become less {\em macroscopically distinguishable }.
However, in the generation of the superposition of two successive Fock states,
decreasing $\omega_m $ really does help. The reason is clear from Eq.(\ref{fsup}). The components of the superposition are Fock states
of the cavity field and thereby do not depend in any way on the parameter $k$. The only contribution of $ k $ comes in the amplitudes, and two successive Fock states can always be made to have significant amplitudes whenever the
measurement outcome $ x $ is about halfway between the peaks of the gaussians $\langle x|\phi_n (\pi)\rangle$ and  $\langle x|\phi_{n+1} (\pi)\rangle$, irrespective of the value of k. Thus there is atleast one type of nonclassical state whose generation can be aided by increasing the mirror frequency.

\section{conclusion}
In conclusion we would like to stress that the work presented here offers new prospects for the observation of nonclassical features of light in the cavity. Even when the decoherence due to the mirror's motion is maximum, the nonclassicality associated with the photon number distribution in the cavity is preserved. However, nonclassicality associated with the phase distribution is hindered. But here also, one can be optimistic from the viewpoint that the effect is not dominant unless the mirror is very heavily damped. We have not addressed the problem of detecting the states produced in the cavity because there exists an extensive literature on this topic~\cite{tom1,tom2,tom3,tom4,tom5,tom6,tom7}.

The primary aim of further work must be to solve the system when the mirror is damped according to quantum Brownian motion models~\cite{point,QN} so that one can set up explicit limits on the parameters such as the temperature and mass of the mirror required to observe the nonclassical states in the cavity in their most undecohered condition. Nevertheless, we expect that the feature of decoherence of the field towards the number state basis should also be present in that full solution, since any physically sensible decoherence model must tend to localize the mirror state towards the coherent state basis. In addition there are prospects to utilize the system for tests of quantum Brownian motion, which can be explored further. Also, we have considered starting initially with only a coherent state inside the cavity. One might be able to produce more interesting nonclassical states by starting with nonclassical states. In fact, all that can be investigated in the Kerr medium can be done with this system. Even more can be done with the mirror system perhaps becuase of the opportunities of conditional measurements.

\section*{Acknowledgments}
This work was supported in part by the UK Engineering and Physical Sciences Research Council, the European Union, the British Council, the Inlaks Foundation and the New Zealand Vice Chancellor's Committee. One of us (PLK) would like to acknowledge  discussions with P.Meystre and S.Stenholm on rubber cavities over
many years.
 
\end{multicols}
\widetext

\begin{appendix}
\begin{multicols}{2}
\section{Derivation of the time evolution operator for the undamped system}
\label{ApA}
The time evolution operator is given by
\begin{equation} 
  U(t) = e^{-i r a^\dagger a} e^{-i t b^\dagger b+ ik t a^\dagger a (b+b^\dagger)}
\end{equation}
where $t$ is the time multiplied by $\omega_m$, $k=g/\omega_m$ and $r= \omega_0/\omega_m$.
We now consider a unitary transformation using the operator
\begin{equation} 
  T = e^{-k a^\dagger a (b^\dagger-b)} .
\end{equation}
Note that this is a displacement operator for the mirror in which the displacement amplitude has been replaced by the number operator for the cavity mode. Using the Baker-Cambell-Hausdorf expansion~\cite{Louisell}, the effect of this transformation is readily shown to be
\begin{eqnarray} 
  T b T^\dagger & = & b + k  a^\dagger a , \\
  T b^\dagger T^\dagger & = & b^\dagger  +  k  a^\dagger a , \\
  T a^\dagger a  T^\dagger & = & a^\dagger a .
\end{eqnarray}
Using the fact that
\begin{equation}
  Uf(\{X_i\})U^\dagger = f(\{UX_iU^\dagger\}), \label{oprel}
\end{equation}
for any function $f$, unitary operator $U$, and arbitrary set of operators $\{X_i\}$, the effect of the transformation on the time evolution operator is easily calculated to be
\begin{equation}
   T U(t) T^\dagger = e^{-ira^\dagger at} e^{-ib^\dagger bt - ik^2 (a^\dagger a)^2 } .
\end{equation}
Multiplying on the left by $T^\dagger$, and on the right by $T$, we obtain the following expression for the time evolution operator:
\begin{equation}
   U(t) = e^{-ira^\dagger at}
          e^{ik^2 (a^\dagger a)^2 t}
          e^{k a^\dagger a (b^\dagger-b)}
          e^{-ib^\dagger bt}
          e^{-k a^\dagger a (b^\dagger-b)}.
\end{equation}
(Note that to obtain this expression we have swapped various exponentials which contain commuting arguments.) To obtain the final form of $U(t)$, we need to swap the last two exponential factors in this expression. To acheive this we note first that the BCH expansion gives
\begin{equation}
    e^{-ib^\dagger b} [a^\dagger a (b^\dagger-b)] e^{ib^\dagger b} = a^\dagger a (b^\dagger e^{-it} - be^{it}) ,
\end{equation}
and again using Eq.(\ref{oprel}) we obtain 
\begin{equation}
    e^{-ib^\dagger b} e^{-k a^\dagger a (b^\dagger-b)} e^{ib^\dagger b} = e^{-k a^\dagger a (b^\dagger e^{-it} - be^{it})} .
\end{equation}
Multiplying both sides on the right by $e^{-ib^\dagger b}$ we arive at the relation required to swap the exponentials, namely
\begin{equation}
    e^{-ib^\dagger b} e^{-k a^\dagger a (b^\dagger-b)} = e^{-k a^\dagger a (b^\dagger e^{-it} - be^{it})} e^{-ib^\dagger b}.
\end{equation}
We may now write the expression for $U(t)$ as
\begin{eqnarray}
   U(t) & = & e^{-ira^\dagger at}
              e^{ik^2 (a^\dagger a)^2 t}
              e^{k a^\dagger a (b^\dagger-b)} \nonumber\\
   & \times & e^{-k a^\dagger a (b^\dagger e^{-it}-be^{it})}
              e^{-ib^\dagger b},
\end{eqnarray}
and to obtain the final expression given in Eq.(\ref{ev}), we need only combine the arguments of the $3^{\mbox{\scriptsize\it rd}}$ and $4^{\mbox{\scriptsize\it th}}$ exponentials, which is readily acheived with the BCH relation~\cite{Louisell}.

\section{Solution to the master equation for the system when the mirror's motion is damped}
\label{ApB}
We need to solve Eq.(\ref{master}), which contains two parts. The solution of the first part
\begin{equation}
  \frac { d\rho (t)}{dt} = -\frac{i}{\hbar} [H,\rho(t)]  ~,
\end{equation}
is known to be
\begin{equation}
\label{unit}
  \rho (t) = U(t) \rho (0) U^\dagger (t) \end{equation}
where, $U(t)$ is given by Eq.(\ref{ev}). On the other hand, the second part,
\begin{equation}
\label{nonunit}
  \frac { d\rho (t)}{dt} = \frac{\gamma}{2} (2 b \rho (t) b^\dagger -  b^\dagger b  \rho (t) - \rho (t) b^\dagger b)
\end{equation} is known to transform $|\lambda_i\rangle \langle \lambda_j |$ as~\cite{Walls2}
\begin{equation}
P_{ij}|\lambda_i\rangle \langle \lambda_j | \rightarrow P_{ij}  \langle \lambda_i |\lambda_j\rangle ^{(1-e^{-\gamma t})}   | \lambda_i e^{-\gamma t/2} \rangle \langle \lambda_j  e^{-\gamma t/2} |
\end{equation}
where, $|\lambda_i\rangle$ and $|\lambda_j\rangle$ are coherent states.

Let the initial state of our system be
\begin{equation}
  \rho (0) = |\alpha \rangle \langle \alpha  |_{\mbox{\scriptsize c}} \otimes |0\rangle \langle 0 |_{\mbox{\scriptsize m}} ,
\end{equation}
that is, we assume the mirror to be in a vacuum state initially (for simplicity). As the damping localizes the mirror state to a coherent state basis, and corresponding to each number state of the field, the mirror is driven to a separate coherent state, at any time $t$ the general form of the density matrix will be
\begin{equation}
\label{break}
  \rho (t) = \sum_{n=0,m=0} \rho_{nm}(t) |n\rangle \langle m |_{\mbox{\scriptsize c}} \otimes |\phi_n (\gamma,t)\rangle \langle \phi_m (\gamma, t) |_{\mbox{\scriptsize m}} .
\end{equation}
Thus, we can split the evolution of the entire system into separate evolutions labelled by $m$ and $n$. In other words, we break our problem into that of calculating the evolution of each of the entities $\rho_{nm}(t) ~ |\phi_n (\gamma,t)\rangle \langle \phi_m (\gamma, t) |_{\mbox{\scriptsize m}}$ separately. In any small time step $\Delta t$ during which evolution occurs only according to Eq.(\ref{unit}), the amplitudes of the coherent states of the mirror change according to
\begin{equation}
  \phi_n (\gamma,t) \rightarrow (1-i \Delta t) \phi_n (\gamma,t) +  ikn\Delta t,
\end{equation}
while if evolution occurs according to Eq.(\ref{nonunit}), then
\begin{equation}
  \phi_n (\gamma,t) \rightarrow  \phi_n (\gamma,t) (1- \gamma \Delta t /2 ) .
\end{equation}
Hence we can set up the following differential equation
\begin{equation} \frac {d \phi_n (\gamma,t) }{dt} = -i \phi_n (\gamma,t)
+ i  k n - \frac{\gamma}{2} \phi_n (\gamma,t) ,
\end{equation}
the solution for which (with initial condition $\phi_n (\gamma,0)=0$) is given by 
\begin{equation}
\label{gphi}
  \phi_n (\gamma,t) = \frac{ikn}{i+\gamma/2 } (1- e^{-(i+\gamma/2)t}) .
\end{equation}
Next let us determine how $\rho_{nm} (t)$ evolves. Following Eq.(\ref{unit}) it evolves as
\begin{equation}
\label{u}
  \rho_{nm} (t) \rightarrow  \rho_{nm} (t) ~ e^{i k^2 (n^2-m^2) \Delta t(1 - \cos{t})} ,
\end{equation}
while Eq.(\ref{nonunit}) transforms
\begin{eqnarray} 
\label{nu}
  \rho_{nm} (t) &  \rightarrow &  \rho_{nm} (t) ~ \langle \phi_n (\gamma,t) | \phi_m (\gamma,t)\rangle ^{(1-e^{-\gamma \Delta t})} \nonumber \\ 
 & = & \rho_{nm} (t)e^{-| \phi_n (\gamma,t) - \phi_m (\gamma,t) |^2 \gamma \Delta t/2 } ,
\end{eqnarray}
where the smallness of $\Delta t$ and the fact that $Im(\phi_n (\gamma,t)  \phi_m (\gamma,t) ^* ) =0$ (from Eq.(\ref{gphi})) have been used. Hence, evolving  $\rho_{nm} (0)$ according to Eq.(\ref{u}) and  Eq.(\ref{nu}) alternately and taking the limit $\Delta t \rightarrow 0$ we obtain
\begin{eqnarray}
\label{ro}
  \rho_{nm} (t) & = &  \rho_{nm} (0) e^{i k^2 (n^2-m^2) (t-\sin{t})}\nonumber \\ 
 & & \times  e^{-\frac{\gamma}{2}\int_0^t | \phi_n (\gamma,t) - \phi_m (\gamma,t) |^2 dt }.
\end{eqnarray}
Combining Eq.(\ref{gphi}), Eq.(\ref{ro}) and Eq.(\ref{break}), and evaluating the integral in Eq.(\ref{ro})  one can obtain Eqs.(\ref{compl1})-(\ref{compl3}) of section~\ref{damp}, which constitutes the complete solution for the density matrix of the system at an arbitrary time.
\end{multicols}
\widetext
\end{appendix}

\end{document}